\documentstyle[12pt,a4,epsf]{article}
\begin{document} 
\bibliographystyle{plain} 
\input{epsf} 
{\Large\bf\noindent Tapping Thermodynamics of the One Dimensional 
Ising Model } 
\vskip 2 truecm
\noindent A. Lef\`evre and D. S. Dean 
\vskip 1 truecm
\noindent 
IRSAMC, Laboratoire de Physique Quantique, Universit\'e Paul Sabatier, 
 118 route de Narbonne, 31062 Toulouse Cedex 04, France.
\pagestyle{empty} 
\vskip 1 truecm \noindent{\bf Abstract:} 
We analyse the steady state regime of a one dimensional Ising model
under a tapping dynamics recently introduced by analogy with the 
dynamics of mechanically perturbed granular media. The idea
that the steady state regime may be described by a flat measure over
metastable states of fixed energy is tested by comparing various steady state
time averaged quantities in extensive numerical simulations 
with the corresponding ensemble averages 
computed analytically with this flat measure. 
The agreement between the two averages is excellent in all the cases examined,
showing that a static approach is capable of predicting certain measurable
properties of the steady state regime.

\vskip 1 truecm
\noindent  February 2001
\newpage
\pagenumbering{arabic} 
\pagestyle{plain}
\section{Introduction}
In complex systems such as granular media the energy available due to thermal
fluctuations is not sufficient to cause particle rearrangement, hence in the
absence of external perturbations the system is trapped in a metastable state.
A granular media may be shaken mechanically and experiments reveal a 
steady state regime defined by an asymptotic compactivity \cite{exps}. 
The non trivial behaviour of these systems, such as slow relaxation dynamics
and hysteresis effects arises from the fact that such systems have an
extensive entropy of metastable or blocked states. A vertically
tapped system of hard spheres tends to a random close packing
\cite{exps}, whereas
a horizontally shaken system crystallises \cite{horiz}, the
stationary states obtained in these systems are not theoretically 
understood for the moment. Edwards \cite{edw} has proposed 
that one may construct 
a thermodynamics over metastable states in the same way as Boltzmann and 
Gibbs  developed over microstates, his hypothesis is that the 
equilibrium measure over these states, in gently tapped systems, is
flat over all blocked states satisfying the relevant macroscopic constraints.
Recently this scenario has been put to the test \cite{jorge} where the authors
examine a form of tapping dynamics on two systems having an extensive entropy
of metastable or blocked configurations, the Kob-Anderson model 
\cite{KA} and the Random 
Field Ising model in three dimensions. They then compared the time average 
quantities obtained in the late time aging regime  with those 
generated by a flat Edwards measure numerically calculated over blocked
configurations of the system. 
While the Edward's measure worked well for the Kob-Anderson model it
was shown to be incorrect for the Random Field Ising model in three
dimensions. Hence, though
not generically true there appear to be systems where the flat measure works.
The 
developement of  a thermodynamics for such systems is important to 
understand the behaviour of granular media, powders and glasses and
has far reaching fundamental  physical and practical applications. Let us 
add that
method of simulated annealing, where an analogy with classical thermodynamics
is used for optimization problems,
 is one of the most robust general methods of 
optimization used in science and industry;
it is possible that tapping type algorithms are more efficient in
certain circumstances (see for example \cite{mon})
and thus the general theory we
are searching for should clearly have applications well beyond physics.

Recently tapping dynamics has been introduced on simple spin systems
in order to draw  analogies with the physics of granular media.
These spin systems were spin glasses and ferromagnets on random thin
graphs \cite{dele1,dele2}
and the three spin model with ferromagnetic coupling on random 
hypergraphs \cite{mehta2}. In the p-spin spherical model a tapping-like
dynamics has been recently introduced using a time-dependent oscillating
magnetic field \cite{berthier}. Simulations on hard sphere systems 
with a similar tapping mechanism were also carried out in
\cite{mehta1}. The general picture emerging from these 
studies and the experiment \cite{exps}
is that the compactivity of the system is increased as the tapping strength
(the amplitude of the external perturbation) is reduced. However
in the horizontal shaking experiments \cite{horiz} this is not the
case. The main reason for choosing the spin systems studied in
\cite{dele1,dele2,mehta2} is that they have an extensive
entropy of metastable states
\cite{mehta2,dean2,lefde}. That is to say that the total number of 
metastable states $N_{MS}$ is given by

\begin{equation}
N_{MS}\sim  \exp(Ns_{Edw})
\end{equation} 

where $s_{Edw}$ is the total Edwards entropy per spin \cite{edw}. The
definition of a metastable state depends on the dynamics of the system, 
here
we will define a metastable state to be a configuration where
no single spin flip reduces the energy of the system. In this paper
we will examine the validity of the
hypothesis of Edwards for the one dimensional
ferromagnet under the tapping dynamics of \cite{dele1,dele2}. If one
implements a zero temperature single spin flip dynamics such that
only single spin flips lowering the energy are permitted, this system has
an extensive Edwards entropy and gets stuck in metastable states.
The dynamics is made to evolve via tapping, that is to say
each spin is flipped in parallel with probability $p \in (0,1/2]$.
The system is tapped after it becomes blocked in a metastable state
and after the tap it relaxes again under the zero temperature or
falling dynamics (by analogy with granular material in a gravitational field).
Here the quantity corresponding to the compactivity is the energy. If one
makes an analogy with granular media, this dynamics corresponds to 
a rapid relaxation dynamics between taps, meaning that the tapping is
characterised only by its strength, $p$, and not by an additional time 
scale. This dynamics is therefore, in a sense, the simplest case of
realistic tapping. The  Glauber dynamics of the one 
dimensional Ising model can be solved analytically \cite{glauber} 
(at all temperatures), 
however the falling dynamics defined here (in between taps) does not seem
amenable to analytic solution. 

The tapping dynamics for this system leads after a sufficiently large
number of taps to a steady state energy $E(p)$ which is constant. This 
value of $E(p)$ can be determined by a mean field theory \cite{dele1,dele2}
which we suspect to be exact as it reproduces certain exact results
obtained combinatorially and also is in perfect agreement with the 
numerical simulations. Here we present numerical simulations under tapping
dynamics to measure quantities such as correlation functions, 
distribution of domain lengths and energy fluctuations (corresponding
to a specific heat) and confront the results with exact calculations
using the Edwards measure.

By the Edwards hypothesis we assume that the stationary measure on 
the tapped systems is
\begin{equation}
\langle O \rangle_{Edw} = {\sum_{{\cal C}} O \exp\left(-\beta(p)
H[{\cal C}]\right)\over Z} \label{edwm}
\end{equation}
where $\{ {\cal C}\}$ is the ensemble of metastable configurations and 
\begin{equation}
Z = \int dE N_{MS}(E)\exp(-\beta(p)NE)
\end{equation}
as suggested in \cite{edw} and recently in
\cite{nico}. 
Here $N_{MS}(E)$ is the number of metastable states with
 energy per spin $E$ and $\beta(p)$ is a Lagrange multiplier 
fixing the energy of the system, which can also be thought of as a
canonical temperature coming from the tapping. Defining the 
Edwards entropy at internal energy $E$, $s_{Edw}(E) = \ln(N_{MS}(E))/N$
we find that the relation determining $\beta(p)$ (the inverse Edwards
temperature) is
\begin{equation} 
\beta(p) = {\partial s_{Edw}(E) \over \partial E}\vert_{E(p)} \label{defb}
\end{equation}
It was shown in \cite{dele1,dele2} that $E(p)$ is a monotonically decreasing
function of $p$ and from the definition (\ref{defb}) and the calculation
of $s_{Edw}(E)$ carried out in \cite{dean2} one may calculate the corresponding
$\beta(p)$ as a function of $p$, which is continuous and monotonically 
decreasing. Hence $p$ plays the  role of an effective temperature
which fixes the internal energy of the steady state.

This does not prove that the Edwards measure is correct for this system
(the hypothesis of Edwards does not tell us how to calculate $\beta(p)$), 
one now has to see if the measure (\ref{edwm}) can be used to compute
quantities in the steady state regime. In this paper we calculate
using the Edwards measure:

\begin{itemize}

\item The internal energy fluctuation per spin 
$c = (\langle {\cal E}^2 \rangle - \langle {\cal E} \rangle^2)/N$ where
${\cal E}$ is the total internal energy, and consequently $\langle {\cal E} \rangle = NE$.
\item The correlation functions $C(r) = \langle S_i S_{i+r}\rangle $
and $D(r) = \langle S_i S_{i+1}  S_{i+r} S_{i+r+1}\rangle$.
\item The distribution of domain sizes.
\end{itemize}
The results are then compared with extensive numerical simulations.

\section{Calculations using the Edwards Measure}

The Hamiltonian we consider is that for the usual one dimensional
ferromagnet with periodic boundary conditions

\begin{equation}
H = -\sum_{i=1}^N S_i S_{i+1}
\end{equation}

The partition function for the system is then given by
\begin{equation}
Z = {\rm Tr} \prod_{i=1}^N \exp(\beta S_{i}S_{i+1}) \theta(S_{i-1}
S_i + S_iS_{i+1})
\end{equation}
where the function $\theta(x) = 0$ for $x<0$ and $\theta(x) = 1$ for 
$x\geq 0$ enforces the metastability (each spin is stable or marginally
stable in its local field). Performing a local change of variables
to new Ising spins $\sigma_i = S_i S_{i+1}$ we obtain
\begin{equation}
Z = {\rm Tr} \prod_{i=1}^N \exp(\beta \sigma_i) \theta(\sigma_i +\sigma_{i+1})
\end{equation}
Hence we  find that $Z = {\rm Tr} \ T^N$ where $T$ is the transfer matrix
\begin{equation}
T = \pmatrix{& a & 1 \cr & 1 & 0}
\end{equation}
with  $a = \exp(\beta)$. From this we find that
\begin{equation}
E = {a\over \sqrt{a^2 + 4}}
\end{equation}
which simply determines $\beta$. The fluctuations in the free
energy are then seen to be given by
\begin{equation}\label{cde}
c = -E(1-E^2)
\end{equation}

The determination of the correlation function $D(r)$ is simple as in
terms of the spins $\sigma_i$, one has  $D(r) = \langle \sigma_{i} \sigma_{i+r}
\rangle$. Using the Edwards measure one obtains
\begin{equation}
D(r) = {{\rm Tr }\ \hat \sigma T^r \hat \sigma T^{N-r} \over Z}
\end{equation}
where $\hat \sigma$ is the matrix
\begin{equation}
\hat \sigma  = \pmatrix{& 1 & 0 \cr & 0 & -1}
\end{equation}
In the large $N$ limit one finds
\begin{equation}
D(r) = \left( \langle t_0 |\hat \sigma | t_0 \rangle \right)^2 + 
\left( {t_1\over t_0}\right)^r
\left( \langle t_1 |\hat \sigma | t_1 \rangle \right)^2 
\end{equation}
where $t_0 = (a + \sqrt{a^2 + 4})/2$ is the maximal 
eigenvalue of $T$ and $t_1 = -1/t_0$ the 
remaining one, $| t_0 \rangle$ and $| t_1 \rangle$ denoting the 
respective normalised eigenvectors. Expressing everything in terms
of the energy per spin $E$ we find that the connected part of $D(r)$ is given
by
\begin{equation}\label{dcder}
D_c(r) = (1- E^2) \left( {E+1\over E-1}\right)^r
\end{equation}

In the same way, one can calculate the two point  correlation function
finding
\begin{equation}
C(r) = C_c(r) = {{\rm Tr}\ (\hat \sigma T)^r T^{N-r} \over Z}
\end{equation}
The matrix $\hat \sigma T$ can be trivially diagonalized: 
$\hat \sigma T=P D P^{-1}$, with
\begin{equation}
P = \pmatrix{u & 1 \cr 1 & u} \ \ \ {\rm and} \ \ 
D = \pmatrix{u & 0 \cr 0 & {1 \over u}}
\end{equation}
with  $u=(a+\sqrt{a^2-4})/2$. This yields
\begin{equation}\label{cder}
C(r) = A\ {\left({u\over t_0}\right)}^r+(1-A)\ {\left({1\over u t_0}\right)}^r
\end{equation}
with
\begin{equation}
A={u^2 t_0^2-1 \over (u^2-1)(t_0^2+1)}
\end{equation}

Defining by $P(r)$ the probability that a given domain has length
 $r$, it is easy to see that this is given by
\begin{equation}
P(r) ={ \langle \delta_{\sigma_1,-1}\delta_{\sigma_2,1}
\cdots \delta_{\sigma_r,1}\delta_{\sigma_{r},1}\delta_{\sigma_{r+1},-1}
\rangle\over \langle \delta_{\sigma_1,-1}\rangle} . \label{eqhier}
\end{equation}
We find that
\begin{eqnarray}
P(r) &=& {1\over t_0^2} \left({a\over t_0}\right)^{r-2} 
\ \ r \geq 2\nonumber \\
&=& 0 \hskip 2.1 truecm r < 2
\end{eqnarray}
Simplifying  this yields
\begin{eqnarray}\label{pder}
P(r) &=& \left({1 + E \over 1-E}\right) 
\left({-2E\over 1-E}\right)^{r-2} = a(E)\exp\left( -b(E)r\right)
\ \ r \geq 2\nonumber \\
&=& 0 \hskip 4.1 truecm r < 2
\end{eqnarray}
Hence, the distribution of domain sizes is geometric for $n>2$, the fact that $P(1) = 0$ is a consequence of metastability as a domain of length $1$ 
would be a single spin surrounded by two antiparallel neighbours which is 
unstable.

\section{Comparison with numerical simulations}

\begin{figure}
\epsfxsize=0.8\hsize
\epsfbox{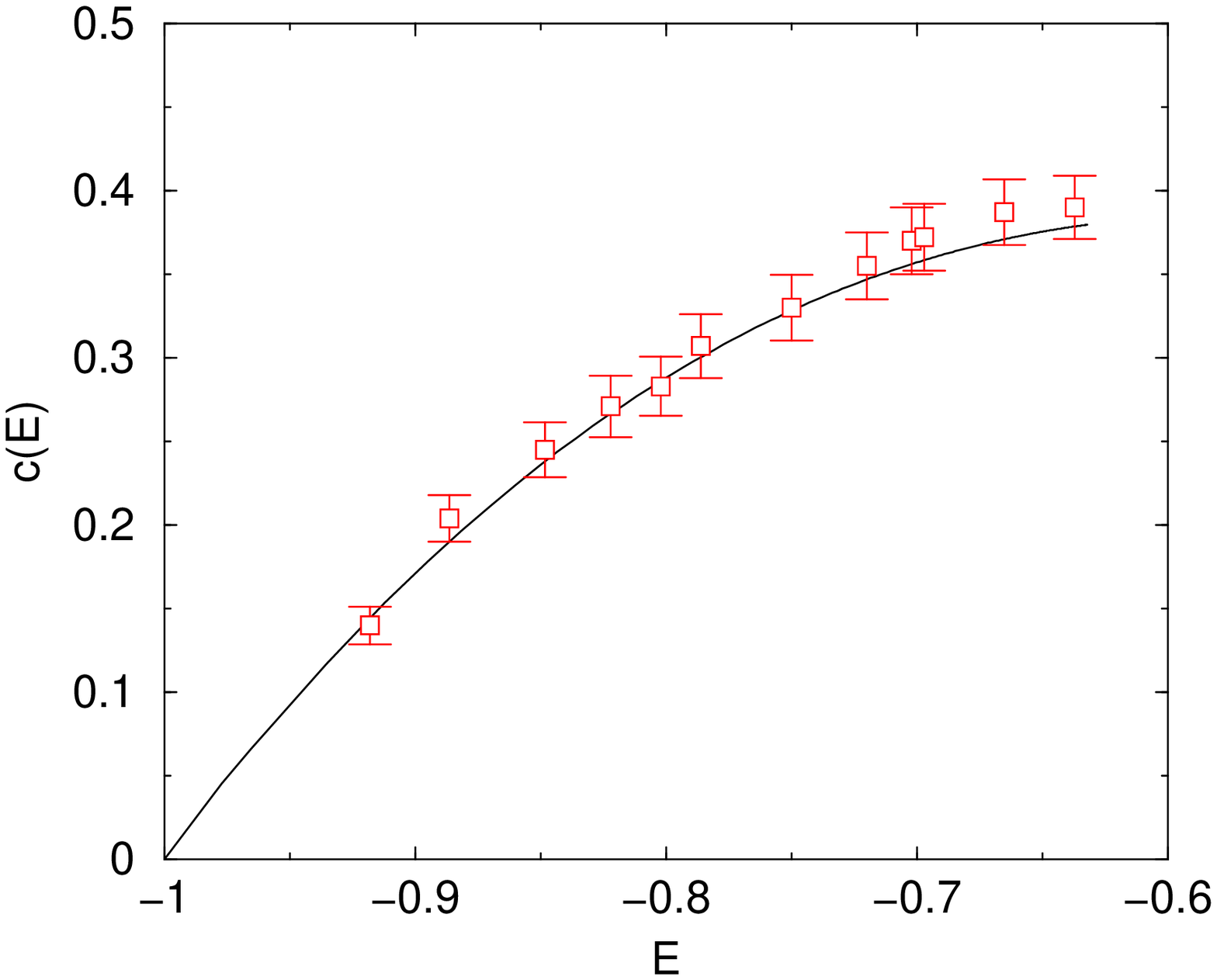}
\caption{The internal energy fluctuation $c(E)$ versus $E$. 
The solid line corresponds to the
 value obtained from Eq. (\ref{cde}) and the symbols are the results obtained
from tapping simulations made on $5000$ systems of $200000$ spins}
\label{fig1}
\end{figure}

\begin{figure}
\epsfxsize=0.8\hsize
\epsfbox{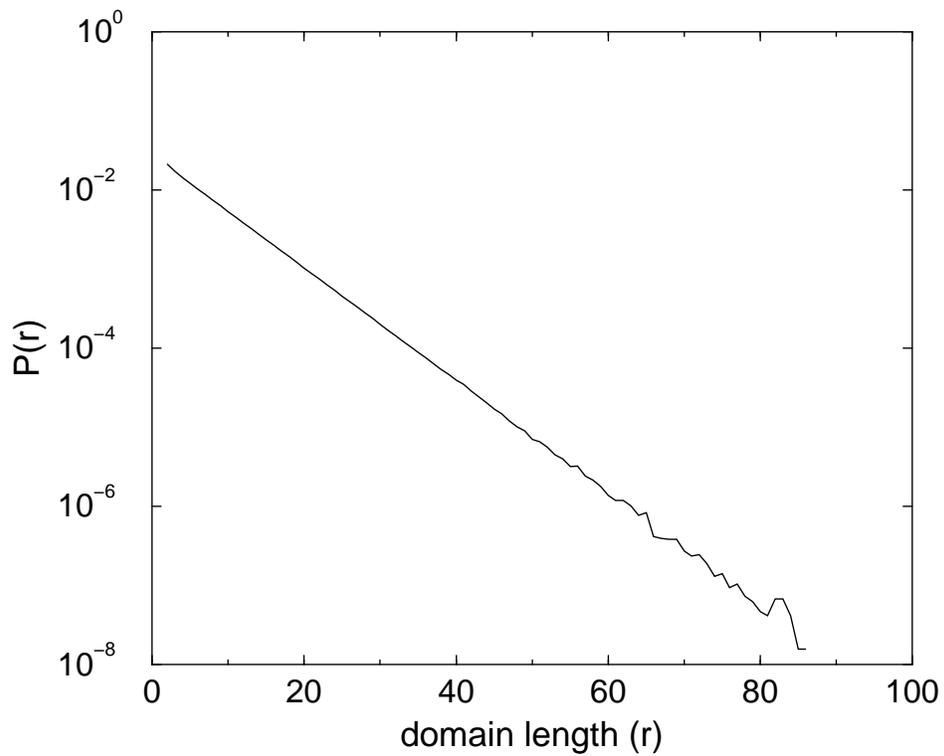}
\caption{Distribution of domain lengths from tapping simulations for
$p=.1$. The vertical scale is logarithmic. The slope is $b=0.165\pm 0.001$,
in excellent agreement with that obtained from Eq. (\ref{pder}).}
\label{fig2}
\end{figure}

\begin{figure}
\epsfxsize=0.8\hsize
\epsfbox{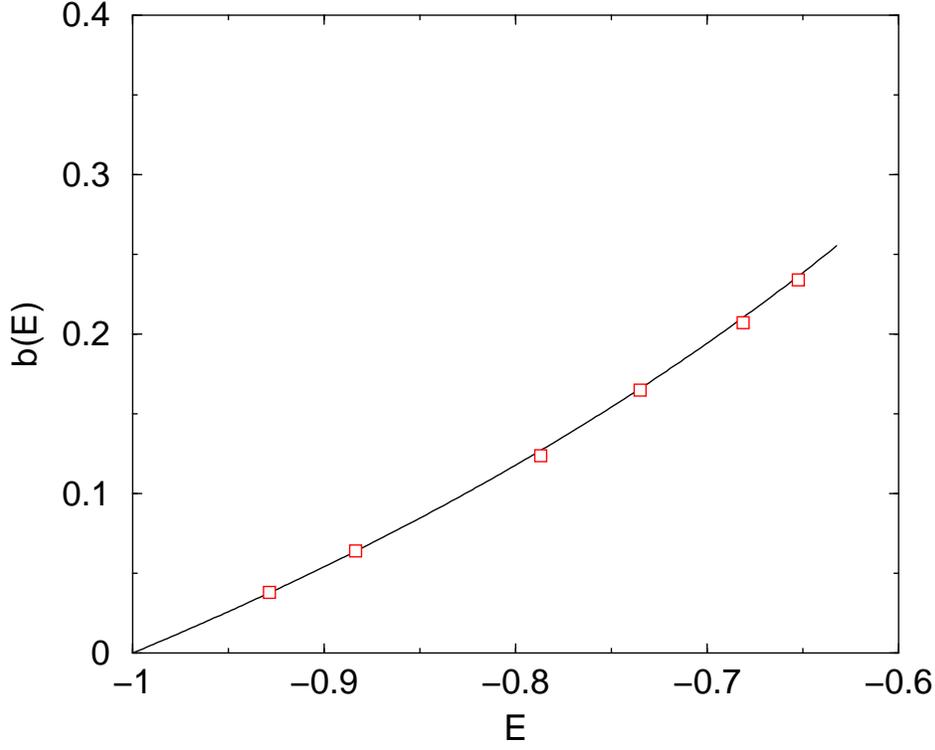}
\caption{Slope of $\ln(P(r))$, $b(E)$ (as defined by Eq. (\ref{pder}))
with respect to the energy. The solid line
correspond to  from Eq. (\ref{pder}) and the symbols
correspond to the same numerics as  Fig.(\ref{fig1}).}
\label{fig3}
\end{figure}

One can compare the results of numerical simulations of tapping with 
the above theoretical ones. For a given value of the energy, let us
say $E$, we have tapped the system with a strength $p$ such that in the
steady state $E=E(p)$. The value of $E(p)$ can be calculated 
\cite{dele1,dele2} and we recall  that $E(p)$ is maximal for $p=1/2$ where
it takes the value $-1+e^{-1}$, and $E(p)$ is monotonically decreasing
for $p\in[0,1/2]$ with  $\lim_{p \to 0^+} E(p) = -1$. The system is tapped
for a sufficiently large number of times , say $t_s$,
to ensure that  the average of the internal energy
$E(t)$ measured becomes stationary. Once in this 
steady state regime, the quantities of interest are measured over a
measurement time (number of taps) $t_{m}=10^5$ . 
The systems  were of size of  
$2\times 10^5$ spins  and 
the results were also averaged over $N_s$ realisations.
Hence, mathematically, the average value of a 
quantity $A$ is calculated, as one would in a Monte Carlo simulation
of a  thermal system, as
\begin{equation}
\langle A \rangle = 
{1\over N_s}\sum_{i=1}^{N_s} {1\over t_m}\sum_{t = t_s + 1}^{t_s+t_{m}} A_i(t)
\end{equation}
In our simulations we found that for the number of sites and $t_m$
used here, that the results obtained from averaging over several systems were
identical to those obtained from measurements over a single system. Hence 
the results are in the thermodynamic limit and the dependence on the system
size vanishes. Consequently the results presented here are from an average over
a single system of size $2\times 10^5$ spins.
In Fig.(\ref{fig1}), we compare the
fluctuation of the energy $c$ calculated using
Edwards' measure, as a function of $E$,  against those   
obtained from the simulations,
the agreement is very good. For small values of $p$ the error bars in our measurements are very small and the agreement with Eq. (\ref{cde}) is 
excellent. For larger values of $p$ the error bars are large as the statistical
fluctuations are larger, however from fig.(\ref{fig1}) we see that the 
value given by Eq. (\ref{cde}) is within the error bars. 
The distribution $P(r)$ of domain sizes, is shown
in fig.(\ref{fig2}) and has a perfect exponential decay
for $r \geq 2$. This guarantees that the result Eq. (\ref{pder}) is in perfect
agreement with the simulations (as the energy $E$ is fixed) as shown in
Fig. (\ref{fig3}). 

\begin{figure}
\begin{minipage}{.52\textwidth}
\epsfxsize=0.9\hsize
\epsfbox{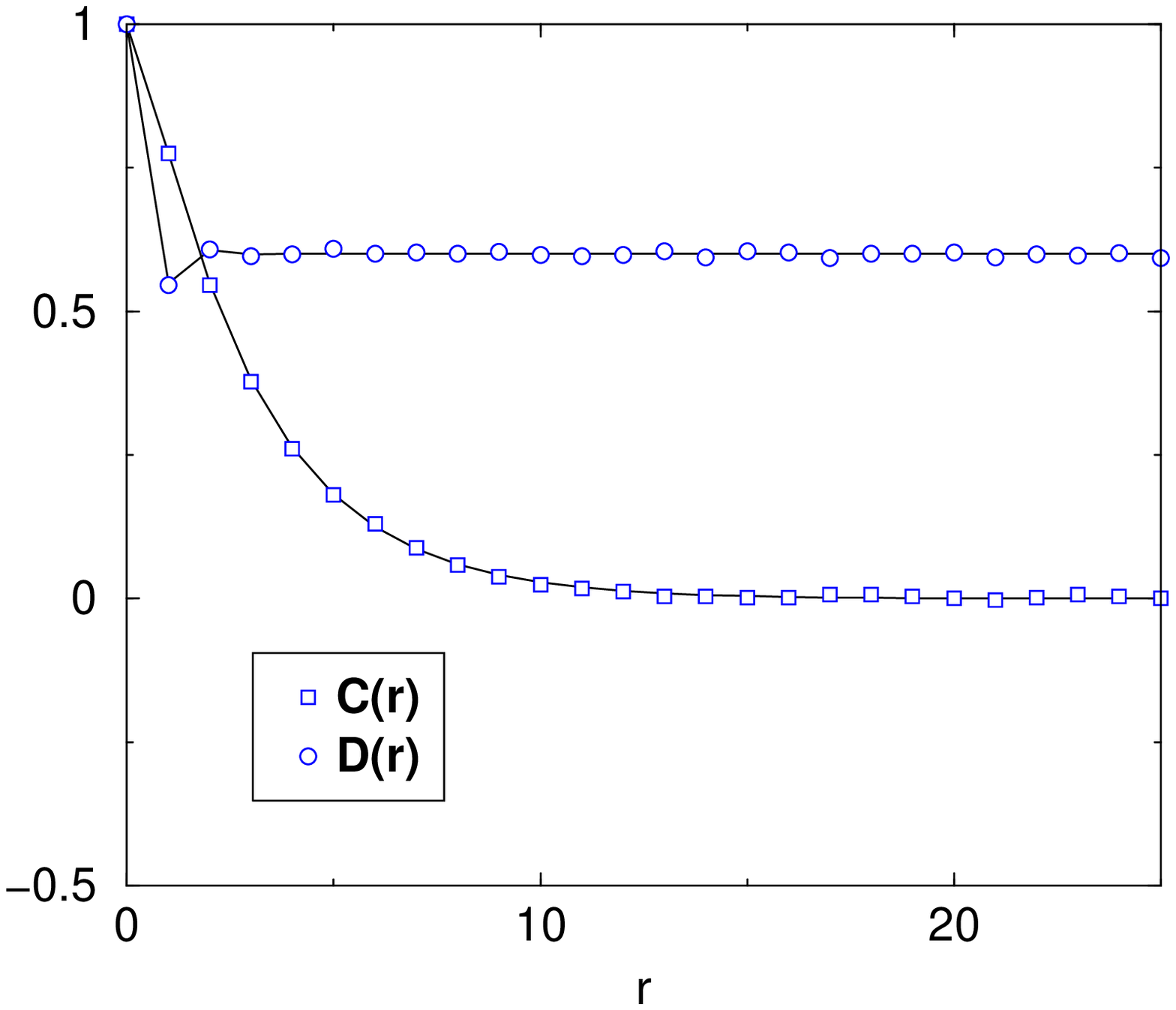}
\end{minipage}
\begin{minipage}{.52\textwidth}
\epsfxsize=0.9\hsize
\epsfbox{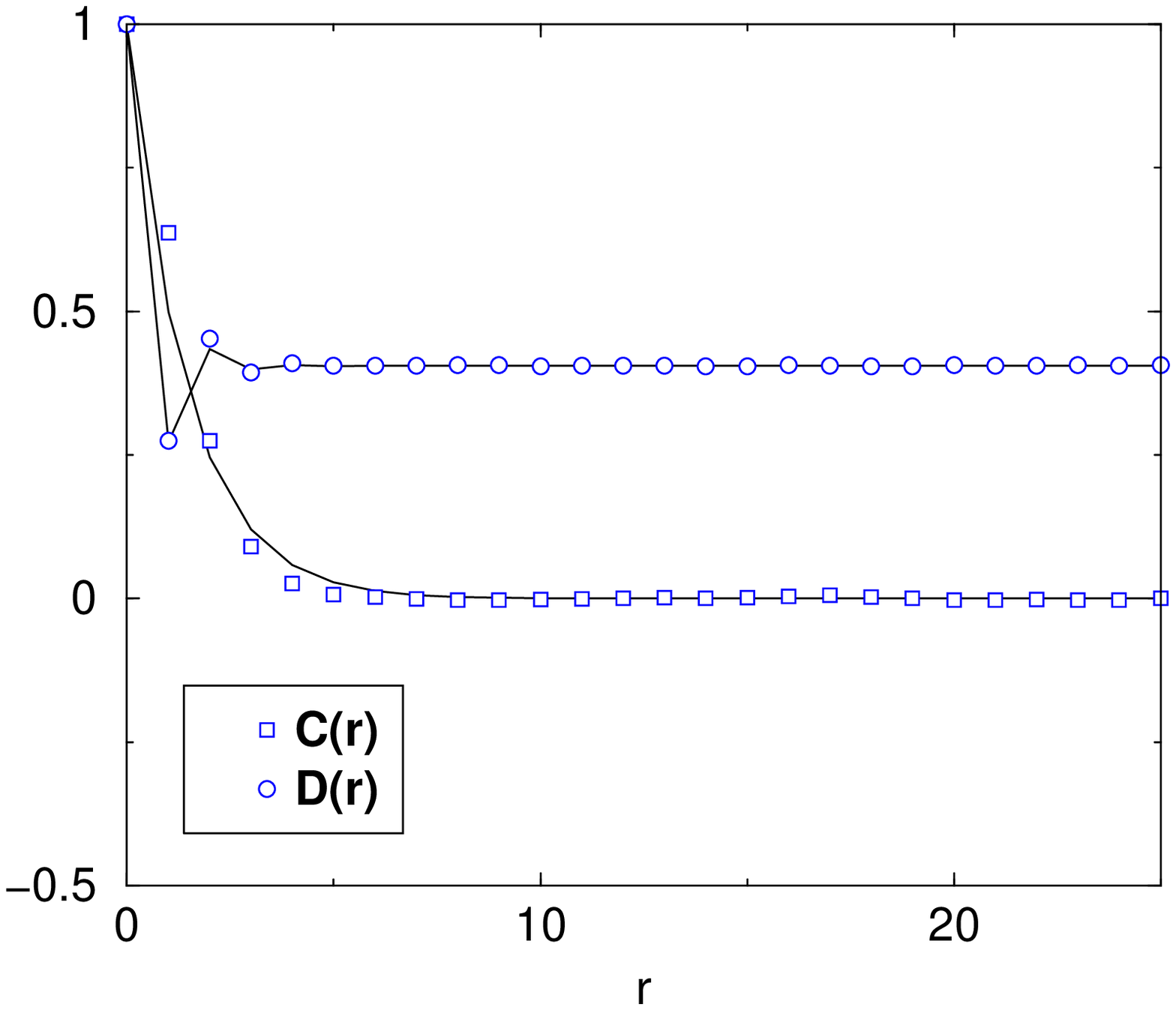}
\end{minipage}
\caption{Comparison between the expected $C(r)$ and $D(r)$ from the
  theoretical calculation with the results from numerical simulations 
for $E=-0.78$ (left) and $E=-0.63$ (right). The symbols are the results of
the tapping experiments and the solid lines correspond to the one predicted by
Eq.(\ref{cder}) and Eq.(\ref{dcder}).}
\label{fig4}
\end{figure}

The correlation functions such $C(r)$ and $D(r)$ have also been computed
numerically. In
Fig.(\ref{fig4}) we have plotted the results 
in comparison with those expected from Eqs. (\ref{dcder}) and (\ref{cder}). 
Here again,
the comparison is excellent (remark that the agreement is better for low
energies, as again the statistical  fluctuations due to the tapping 
are much smaller for low $p$ than for high $p$). 

\section{Conclusion}
We have simulated numerically
tapping dynamics on a one dimensional system and measured
fluctuations of the energy, correlation functions and distributions of
domain sizes. The values of
these quantities expected from a thermodynamics built by using a flat
measure over blocked configurations agrees very well with our 
simulation data. The use of the flat measure therefore allows one to 
accurately predict the two point correlation function, a particular
four point function  $D(r)$ and also a hierarchy of 
$n$ point functions 
(see Eq. (\ref{eqhier}))
corresponding to the distribution of domain sizes $P(r)$.
In principal further quantities could be investigated but the numerical
study would be hampered by statistical fluctuations. The important 
point here is that the principal quantities open to experimental determination
are well predicted from Edwards' measure. A proof of the absolute validity of
the use of the flat measure seems difficult, there is no obvious form
of detailed balance in the tapping dynamics and one would need to show that
at fixed internal energy per spin the system explores uniformly the metastable
states of this energy. Physically this seems quite likely in the limit of
small tapping. Here a metastable state can be viewed as a configuration of
islolated domain walls. The first order $O(p)$ effect of tapping creates 
domain wall diffusion and annihilation upon the encounter of two domain walls
as in the zero temperature Glauber dynamics of the Ising model
\cite{glauber,bray}. In addition within a domain the flipping of two
consecutive spins creates a domain of length $2$ which then con tributes the 
the diffusion/annihilation process mentioned previously, this process is
however  $O(p^2)$. Physically, the steady state is then reached upon the 
equilibrium between the creation of small domains of this type within larger
domains and the annihilation of domains driven by diffusion. Hence, as the
steady state regime is characterised by an average number of domain walls
with the annihilation and creation processes mentioned above in equilibrium,
it seems plausible the diffusion generated by the $O(p)$ tapping enables the
system to explore the configurations available in a flat manner.

{\bf Acknowledgments:} During this work we have benefitted 
from useful and illuminating discussions
with Jorge Kurchan, Satya Majumdar and Erik Sorensen.

\newpage

\end{document}